\documentclass[american,aps,reprint, superscriptaddress]{revtex4-1}
\usepackage[unicode=true,pdfusetitle, bookmarks=true,bookmarksnumbered=false,bookmarksopen=false, breaklinks=false,pdfborder={0 0 0},backref=false,colorlinks=false] {hyperref}
\hypersetup{ colorlinks,linkcolor=myurlcolor,citecolor=myurlcolor,urlcolor=myurlcolor}
\usepackage{graphics,epstopdf,graphicx, amsthm, amsmath, amssymb, times, braket, colortbl, color, bm, soul}
\usepackage[up]{subfigure}
\usepackage{cleveref}

\definecolor{myurlcolor}{rgb}{0,0,0.7}

\theoremstyle{plain}

\providecommand{\theoremname}{Theorem}

\newcommand*{\myproofname}{Proof}

\makeatother

\begin{document}

%\title{Almost Markovian maps and corresponding non-Markovoianity measures, and bound on the latter based on system-environment entanglement}
%\title{Almost Markovian maps and entanglement-based bound on corresponding non-Markovianity measure}
\title{Nearly Markovian maps and entanglement-based bound on corresponding non-Markovianity}
\author{Sreetama Das\(^1\), Sudipto Singha Roy\(^{1,2}\), Samyadeb Bhattacharya\(^{1,3,4}\),
% Aditi Sen(De)\(^1\), 
Ujjwal Sen}
\affiliation{Harish-Chandra Research Institute, HBNI, Chhatnag Road, Allahabad 211 019, India\\
\(^2\)Instituto de F{\'i}sica T{\'e}orica UAM/CSIC, Madrid, Spain
    \\
\(^3\)S. N. Bose National Centre for Basic Sciences, Salt Lake, Kolkata 700 098, India
\\
\(^4\) Center for Security Theory and Algorithmic Research, International Institute of Information Technology, Gachibowli, Hyderabad, India}
\begin{abstract}
We identify a set of dynamical maps of open quantum system, and refer to them as ``$ \epsilon $-Markovian" maps. It is constituted of maps which, in a higher dimensional system-environment Hilbert space, possibly violate Born approximation but only a ``little". We characterize the ``$ \epsilon$-nonmarkovianity" of a general dynamical map by the minimum distance of that map from the set of $ \epsilon $-Markovian maps. We analytically derive an inequality which gives a bound on the $ \epsilon$-nonmarkovianity of the dynamical map, in terms of an entanglement-like resource generated between the system and its ``immediate'' environment. In the special case of a vanishing \(\epsilon\), this inequality gives a relation between the $\epsilon$-nonmarkovianity of the reduced dynamical map on the system and the entanglement generated between the  system and its immediate environment. We numerically investigate the behavior of the similar distant based measures of non-Markovianity for classes of amplitude damping and phase damping channels.
\end{abstract}
\maketitle

\section{Introduction}

The study of open quantum systems is of fundamental importance in several areas, including the field of quantum information. In an ideal scenario, the evolution of a closed quantum system is described by a unitary operation and is mathematically described by the Schr{\"o}dinger's equation. But in the real world, a system is never perfectly isolated. The interaction with the environment gives rise to non-unitary evolution of the quantum system which causes dissipation of energy and loss of coherence. In arguably the simplest case, the mathematical model of the evolution of an open system is derived using a number of assumptions, which, collectively, has been christened as the ``Markovian'' approximation \cite{petru}. A principal underlying assumption is that the coupling between the system and the environment is weak, so that the environmental excitations decay in a time much shorter than the time it takes for the system to evolve from initial state.
%the flow of information between the system and environment is uni-directional; 
For such a process, the information that flows from the system to the environment can never come back to the system, i.e. the system does not have a memory. In contrast to that, if the interaction between system and environment is such that the information flow can happen both ways,  the system is said to have memory-effects. Several variations of this conceptualization of quantum Markovianity and non-Markovianity exists in the literature
%, and Such an evolution is referred to as non-Markovian 
for evolution of open quantum systems \cite{petru, other-than-Petru, TnaDbaro, other1}. 

Over the years, a number of non-Markovianity measures have been proposed. 
%which may be classified into two broad categories. 
This includes a measure that identifies non-Markovianity by studying 
the time-dynamics of entanglement between the system and an auxiliary system \cite{RHP}. A Markovian evolution causes monotonic decrease of the entanglement, whereas a non-Markovian evolution may give rise to consecutive decay and revival of the entanglement between system and auxiliary system. This behavior of entanglement can be captured in the divisibility property of the dynamical map of the reduced system. 
%If the dynamical map is divisible and forms a complete positive (CP) semigroup, the evolution of the system is Markovian, and if not then the evolution is nonmarkovian \cite{wolf}. 
Using the concept of divisibility of dynamical maps \cite{wolf}, Rivas \emph{et al.}
%(RHP)
 \cite{RHP} formulated a necessary and sufficient criterion to detect non-Markovianity when the exact form of the dynamical map of the reduced system is known. 
%This class of measures, commonly called as RHP (Rivas-Huelga-Plenio) measure, has been 
Further studies in this direction include \cite{odvut}. 
There are several works that look for manifestation of  non-Markovianity in the non-monotonic behaviour of a number of other quantum-mechanical properties of a system, e.g. flow of quantum Fisher information \cite{lu}, fidelity difference \cite{usha}, quantum mutual information \cite{luo}, volume of accessible states of a system \cite{lorenzo}, accessible information \cite{fanchini}, total entropy production \cite{salimi}, quantum interferometric power \cite{dhar}, coherence \cite{chanda}, etc. 
Another class of measures, proposed by Breuer \emph{et al.}
%(BLP)
 \cite{BLP} (see also \cite{rattir-baroTar-par-Kolkata-shasan-kare-charjan-jubak}),
% and broadly known as BLP (Breuer-Laine-Pillo) measure, 
associates the distinguishability of quantum states with the non-Markovian behavior of their evolution. A backflow of information from the environment to the system possibly increases the distinguishability, whereas in case of Markovian evolution, the one-way information flow from the system to the surroundings results in monotonic decrease of the distinguishability of the quantum states \cite{BLP1, MLBMP, AFPP}. However, there are instances when these different non-Markovianity measures are not in agreement with each other. Specific examples show that the evolution of an open quantum system can be Markovian according to BLP but the corresponding dynamical map is indivisible and hence the evolution is non-Markovian according to RHP \cite{haikka,CMR}. Another work demonstrates that the BLP measure is not equivalent to a non-Markovianity measure based on correlations \cite{lu, jiang}. Though there has been attempts to correlate the different measures,  a clear understanding and quantification of non-Markovianity and its relatively subtler issues have remained elusive (see Refs. \cite{haikka, CMR, jiang, Modi_etal} in this regard). There exist also quite recent works on comparison of the efficiency and hierarchy of non-Markovianity measures
%in more complicated qubit dynamics 
and the hierarchy between the definitions of quantum Markovianity in the general level without specifying the type of the dynamical map \cite{TnaDbaro, ei-pathe-eka-eka-hnaTten}.
%We need a measure which will completely characterize the non-Markovianity of a quantum system.

In our work, we propose a distance-based measure of non-Markovianity which is independent of the above two characterizations (see \cite{RHP} in this regard). With the usual picture of a system and its environment, we consider an additional bath (environment), which is much larger than the environment immediate to the system, and in which our system and environment are immersed. A set of maps, called 
$\epsilon$-Markovian maps, are conceptualized, and $\epsilon$-nonmarkovianity of a dynamical map is defined as the minimized distance of that map from the set of $ \epsilon$-Markovian maps. We derive an inequality which gives a bound on the above measure of non-Markovianity of a general dynamical map, in terms of an entanglement-like quantity. In the special case of $ \epsilon=0$, we obtain this bound on non-Markovianity in terms of an entanglement \cite{HHHH} of the system-environment joint state. To get an idea how the optimized distance based measures of non-Markovianity behave temporally, we numerically study the behavior of such distance functions for an amplitude damping channel and a phase damping channel, where, depending on the range of the respective parameters, the channels can behave as a Markovian or a non-Markovian map. We note here that the actual evaluation of measure proposed here is a difficult optimization problem, and even the numerical procedures are pursued with some assumptions (which we clearly mention). We believe however that despite the seemingly difficult optimization process making the evaluation of the measure difficult, it will be a useful tool in a better understanding of the concept of Markovianity, its absence, and a quantification of its absence. 

The remainder of the paper is arranged as follows. In Section \ref{dui}, we briefly summarize the concept of quantum mutual information, as it will be the main information-theoretic tool for the formulation of the concept of Markovianity and its absence in this paper. In Section \ref{tin}, we formally define the concept of \(\epsilon\)-Markovianity and Markovian-like maps, and their complementary sets. We first briefly review the classical case, and then move over to the quantum one. We also provide an example of an \(\epsilon\)-Markovian map. Section \ref{char} contains the definition of non-Markovianity and \(\epsilon\)-nonmarkovianity. Section \ref{pnach} contains the numerical simulations for evaluations of distance based non-Markovianity measure for the amplitude and phase damping channels. The entanglement-based bound on non-Markovianity (in the \(\epsilon = 0\) case) is presented in Section \ref{chhoi}, with the bound in the general case being given in the succeeding section (Section \ref{saat}). Section \ref{saat} also contains the definitions of \(\epsilon\)-separability and \(\epsilon\)-entanglement. The measure is defined by using min-distance in Section \ref{aaT}, and the Choi-Jamio{\l}kowski-Kraus-Sudarshan approach in Section \ref{noi}. A conclusion is given in Section \ref{dos}.

%\section{A relative measure} 

\section{Quantum mutual information}
\label{dui}

We now take a short paragraph to discuss a few aspects of the concept of quantum mutual information, as it is central to the definition of non-Markovianity in this paper. The classical mutual information \cite{qmi-1} between two random variables \(X\) and \(Y\) is defined as \(I_c(X:Y) = H(X) +H(Y) - H(X,Y)\), where \(H(X)\) and \(H(Y)\) are the Shannon entropies of \(X\) and \(Y\) respectively. \(H(X,Y)\) is the Shannon entropy of the joint probability distribution of \(X\) and \(Y\). The quantum mutual information \cite{qmi-2, qmi-3} can be seen as the simply replacing the Shannon entopies by von Neumann ones \cite{wehrl} for the local states and the global two-party state. However, the classical mutual information can be expressed in more than one way, and in fact the difference between the quantum generalizations of two such forms led to the conceptualization of ``quantum discord'' \cite{bera, discord-papers}. But quantum mutual information can also be defined as the relative entropy distance  \cite{wehrl} of the two-party quantum state from the tensor product of its local states. This has led to the identification of the quantum mutual information as the total correlation present in the corresponding two-party quantum state. We must mention here that relative entropy is not a true ``distance" measure according to strict mathematical definition, since it is not symmetric with respect to its arguments. Another way to understand the same identification is given in Ref. \cite{qmi-3}, where they find that the amount of noise needed to destroy all correlations present in a two-party state is exactly given by its quantum mutual information content. 

\section{Markovian-like and \(\epsilon\)-Markovian maps, and their complementary sets}
\label{tin}

\subsection{The classical case, briefly}
\label{tin-ka}

The concept of Markovianity is an important and well-known one in the dynamics of systems governed by classical mechanics \cite{Swarnendu, other1}.  A classical stochastic process is a random variable \(X\) that depends on a parameter, say, \(t\). The parameter \(t\) can be understood to indicate the number of steps (in time). Such a process is called Markovian if the value that the random variable assumes at a certain time-step is dependent only on that of the previous step. If at the \(n\)th step, the parameter \(t=t_n\) and \(X(t_n)= x_n\), then the conditional probability of \(X=x_n\) at \(t=t_n\) given that \(X=x_i\) at \(t=t_i\), for \(i<n\) is the same as the conditional probability of 
\(X=x_n\) at \(t=t_n\) given that \(X=x_{n-1}\) at \(t=t_{n-1}\), for all \(x_i\) and \(t_i\) with \(i=1, 2, \ldots, n\). It is this classical notion of Markovianity vis-{\`a}-vis its absence that we wish to generalize to the quantum case.
For the generalization, it will be important to note that for Markovian stochastic matrices, \(T(x_n, t_n \leftarrow x_m, t_m)\), connecting probabilities at times \(t_n\) and \(t_m\), with \(n > m\), we have the following ``divisibility'' property: \(T(x_3, t_3 \leftarrow x_1, t_1) = \sum_{x_2} T(x_3, t_3 \leftarrow x_2, t_2) T(x_2, t_2 \leftarrow x_1, t_1)\), for all \(x_1, x_2, x_3\) and for all \(t_1 \leq t_2 \leq t_3\). It is a Chapman-Kolmogorov relation for Markovian stochastic matrices, and says that the evolution can effectively be considered as having been realized in several intermediate steps. It is also to be remembered that divisible classical processes can be non-Markovian.

%\section{Our model}

\subsection{The quantum version}
\label{tin-kha}

\begin{figure}[t]
\includegraphics[width=0.25\textwidth]{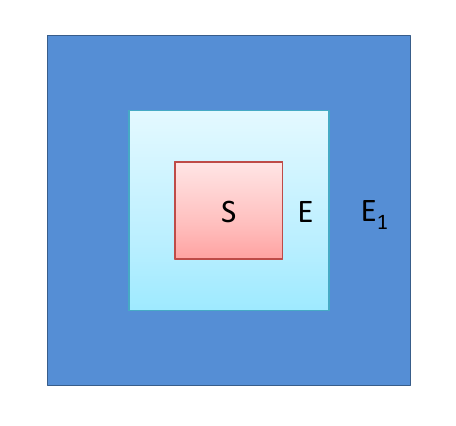}
\caption{(Color online.) A schematic diagram showing the system $S$ immersed in environment $E$. The joint system $ SE $ is in contact with a larger environment $ E_{1} $.}
\label{schematic1}
\end{figure}

We begin by considering a quantum system $ S $ in contact with an environment $ E $. The joint system $ SE $ is immersed in a much larger environment $ E_{1} $. See Fig. \ref{schematic1}. The corresponding Hilbert spaces are denoted by $ \mathcal{H}_{S} $, $ \mathcal{H}_{E} $, and $ \mathcal{H}_{E_1} $. Initially,  the total system $ SEE_{1} $ is a product of three states of the three subsystems, \(S\), \(E\), and \(E_1\). 
%state along the $ S:EE_{1} $ bipartition, so that 
The reduced system $ SE $ is thereby also a product state along the $ S:E $ partition.  As time goes by, the total system evolves unitarily and becomes entangled across different partitions. The reason that we consider a larger environment $ E_{1} $ in which the system-environment duo $ SE $ is immersed will become clear later when we discuss about Markovian-like maps. If we look at the reduced system $ SE $, the time evolution can be described by the dynamical map $ \Lambda_{SE} $. Thus, at any time $ t $, the state of $ SE $ is \begin{equation}
\rho^{SE}(t) = \Lambda_{SE}(\rho^{S}_{0} \otimes \rho^{E}_{0}), 
\label{one}
\end{equation}
where $ \rho^{S}_{0} $ and $ \rho_{E}^{0} $ are the initial states of $ S $ and $ E $.
It may be noted here that if we assume that the initial state of \(SEE_1\) is a product across \(S:EE_1\) only, allowing correlations (classical or quantum) between \(E\) and \(E_1\), the map 
\(\Lambda_{SE}\) will depend on the state of the entire \(SEE_1\), and it may in general be not completely positive \cite{jater-phatna}. 

The division between the environments \(E\) and \(E_1\) has been defined as follows. \(E\) is that part of the environment that directly interacts with the system, while \(E_1\) is the one that is required to take \(E\) on a “thermalization” path before  the next time-step in its evolution with the system. Strictly speaking, this division is arbitrary, although we believe that in physically interesting cases, this division will be a natural one. We will come back to such points later in the paper. For example, in case of a superconducting qubit, the phonons that come in immediate contact of the qubit can be considered as the environment $E$. All the other vibrational modes which do not interact with the qubit, constitutes the environment $E_{1}$. 

Let us consider a particular subset of dynamical maps $ \tilde{\Lambda}_{SE}^{\epsilon} $ such that that for any fixed $ \epsilon \geq 0$, the time-evolved state $ \tilde{\rho}^{SE}(t) $ satisfies the inequality
\begin{equation}
I_{Q}(\tilde{\rho}^{SE}(t)) \leq \epsilon, \hspace{1cm}	\forall \hspace{0.1cm} t, \rho^{S}_{0},\rho^{E}_{0},
\label{two}
\end{equation}
where $ I_{Q}(\varrho^{AB}) = S(\varrho^{A}) + S(\varrho^{B}) - S(\varrho^{AB}) $ is the quantum mutual information \cite{qmi-1, qmi-2, qmi-3}, of a bipartite state $ \varrho^{AB} $, whereas $ \varrho^{A} $ and $ \varrho^{B} $ are respectively  the reduced states of subsystems $ A $ and $ B $, and $ S(\cdot) $ is the von Neumann entropy \cite{wehrl} of its argument.

Let us  mention here that in Ref. \cite{mairi-kono-connection-nei-kintu},
%We thank the Referee for bringing the reference B. Bylicka, D. Chrusciński, and S. Maniscalco. Scientific Reports, 4 5720 (2014) to our attention. 
the authors use the non-monotonicity in time-evolution of a classical and a quantum capacity of a map to define the non-Markovianity of that map. The classical capacity is defined in terms of quantum mutual information between the input and output of the channel. This is different from our approach, as we use the quantum mutual information in the system-environment output state of the evolution of the system-environment duo. The quantum capacity is defined by using a coherent information between the input and output of the map.

The quantum mutual information is a non-negative quantity, so that 
\(I_{Q}(\tilde{\rho}^{SE}(t))\) in inequality (\ref{two}) is lower bounded by zero.
We will refer to the corresponding reduced maps $ \tilde{\Lambda}_{S}^{\epsilon} $ of system $ S $ as $ \epsilon $-Markovian. The set of all such $ \epsilon $-Markovian maps is denoted by $ S^{\epsilon} $.
%%%%%%%%%%%%%%%%%%%
%%%%%%%%%%%%%%%%%%%%
Therefore, for the vanishing \(\epsilon\) case, 
at every instant of time, the state of the system \(S\) and the environment \(E\) is a product state. However, the dynamical map $\Lambda_{SE}$ is not unitary in general. Hence, even in the case $\epsilon=0$, the dynamics of $S$ is not complete positive. This does not affect the classification of ``$\epsilon$-Markovian" maps, since it does not depend on the definition of Markovianity in terms of divisibility. 
%so that a unitary dynamics on \(SE\) starting off from that instant will lead to a  completely positive map on \(S\). 
%This possibility of having a completely positive map on \(S\) at every step of the evolution for arbitrary unitaries on \(SE\) acting at arbitrary times, is considered here as the parallel in the quantum case of the Chapman-Kolmogorov relation for transition matrices in the classical case. 
%Since such maps in the classical case can also be non-Markovian, 
We will call the reduced dynamical maps on $S$ in the case $\epsilon=0$ as ``Markovian-like'' instead of ``Markovian''. More about these maps has been discussed in Section \ref{pnach}.
%
%we have constrained our non-Markovian maps to 
%lie within the set of non-divisible maps. 
See \cite{Kavan-ebang-Samya} in this regard, where Markovian maps are defined by using the concept of a certain quantum process tensor.

It is to be noted that, a number of previous works have investigated non-Markovian dynamics  by dividing the actual infinite dimensional environment into two parts \cite{pseudomode1, embedding, reaction-coordinate, NMV-core}. One of them consists of finite number of degrees of freedom of the environment which significantly interacts with the system, the other is the residual infinite dimensional part. This separation leads to considerable mathematical simplicity while taking into account environmental interactions on the system dynamics, and often shows good agreement with experimental results. Depending on the crude details of identifying relevant environmental subsystems, the several approaches that has been proposed are Markovian-embedding \cite{embedding}, pseudomode method \cite{pseudomode1}, reaction coordinate model \cite{reaction-coordinate}, non-Markovian core model \cite{NMV-core} etc. What remains common in these works, is that the non-Markovian dynamics of the system is attributed to the effective environment, whereas they together are treated with Born-markov approximation with respect to the residual environment. Unlike this, we do not consider a Markovian approximation between joint system $SE$ and $E_{1}$. Ours is a more general picture in which the correlation generated between $SE$ and $E_{1}$ depends on the dimension of $SE$, hence a Markovian approximation will be a very particular specialization.

\subsection{Building an \(\epsilon\)-Markovian map}
\label{tin-ga}

Let us consider the above definition in more detail. 
%We will call the \(\epsilon\)-Markovian maps for \(\epsilon =0\) as Markovian-like maps. 
Markovian-like maps on \(S\) are therefore those, which for any given product initial state between \(S\), \(E\), and \(E_1\), does not create any correlation (as quantified by the quantum mutual information) between \(S\) and \(E\). This parallels the Born-Markov approximation in the ``derivation'' of the dynamics of open quantum systems, where at every time-step, that is much shorter than the typical time required for a nontrivial interaction between the system and its environment, the system is brought back to its initial state and all correlations between the system and the environment are removed. 
The \(\epsilon\)-Markovian maps are introduced in view of the fact that while almost no naturally occurring physical map is Markovian-like, there may be members of important families of maps that violate the criterion for being Markovian-like only ``slightly'', in the sense of creating only a small amount of quantum mutual information for all time. Let us immediately provide an example of such a map, albeit an artificial one. Consider a qubit \(S\) interacting with an ``environment'' \(E\) which is also a qubit, and a bigger environment \(E_1\) consisting of a large number of qubits.  Initially, the state of the entire system is \(|0\rangle_S |0\rangle_E |00 \ldots \rangle_{E_1}\). A unitary \(U_{SE}(1)\) acts on \(SE\) to transform the state of \(SE\) to 
\(|\psi\rangle_{SE}=\alpha |00\rangle + \sqrt{1-|\alpha|^2}|11\rangle\), where \(\alpha\) is a real number such that \(I_Q(|\psi\rangle_{SE}) = \epsilon\), for a previously chosen, possibly small, fixed real number \(\epsilon \in [0,2]\). The joint system is four dimensional, hence defining the unitary on one particular basis will give rise to an infinite number of possibilities to define its action on the remaining three basis vectors. To unambiguously define the unitary, we will need to settle on its action on the remaining three basis vectors. We assume that this has been done. 
%is of course not unambiguously defined 
After this step, a swap operator acts on the qubit \(E\) and a single qubit of \(E_1\), so that the whole system is now 
\begin{equation}
|\psi\rangle_{SE_1^{r}} |0\rangle_E |00 \ldots \rangle_{E_1^{\bar{r}}},
\end{equation}
where the superscript \(r\) denotes a qubit of the environment \(E_1\), while the superscript \(\bar{r}\) denotes its complement. 
This entire evolution, viz. the action of the unitary \(U_{SE}(1)\) and the swap operator, is to be considered as a single time-step. The state of \(SE\) after this time step is \((|\alpha|^2 |0\rangle \langle 0| + (1-|\alpha|^2) |1\rangle \langle 1|)_S \otimes (|0\rangle \langle 0|)_E\). In the next time-step, we act with \(U_{SE}(\epsilon_1) = \exp{(i\epsilon_1 H_{SE})}\), to create the state \(\rho_{SE}\), where 
\(U_{SE}(1) = \exp{(i H_{SE}})\), and where \(\epsilon_1\) is chosen so that \(I_Q(\rho_{SE}) \leq  \epsilon\). After this operation and still remaining within the same time-step, we swap the state of \(E\) with the state of a qubit of \(E_1\) that is still in \(|0\rangle\). This second time-step is repeated again and again. We are assuming that \(E_1\) has enough qubits to do these repetitions. We find that the construction of the dynamics is such that the quantum mutual information is never more than \(\epsilon\), and thus, the above definition leads us to call it an \(\epsilon\)-Markovian map. In the entire construction of the map, we did not make any assumption that required us to go beyond quantum mechanics.   
We will discuss more about these issues later in the paper.

\section{Measure of non-Markovianity and \(\epsilon\)-nonmarkovianity}
\label{char}

It is to be noted that it is the amount of violation of \(\epsilon\)-Markovianity that we will refer as 
\(\epsilon\)-nonmarkovianity, and so it is really non-``\(\epsilon\)-Markovianity'' that we will be dealing with. But we prefer to call it as \(\epsilon\)-nonmarkovianity to avoid a cumbersome nomenclature. 

Our goal now is to quantify the non-Markovianity of a general dynamical map $ \Lambda_{S} $ by its distance $ \mathcal{D} $ from the $ \epsilon $-Markovian maps $ \tilde{\Lambda}_{S}^{\epsilon} $, minimized over the set $ S^{\epsilon} $. We call it $ \epsilon$-\textit{nonmarkovianity}  of the corresponding map $ \Lambda_{S} $, and denote it by $ N^{\epsilon}(\Lambda_{S}) $. That is,
\begin{equation}
    N^{\epsilon}\big(\Lambda_{S}(t)\big) =\min_{\tilde{\Lambda}_{S}^{\epsilon} \in S^{\large \epsilon}}  \mathcal{D}\big(\Lambda_{S}(t)||\tilde{\Lambda}_{S}^{\epsilon}(t)\big).
    \label{three}
\end{equation}

The distance $ \mathcal{D} $ on the space of maps can be conceptualized in a variety of ways. Later on, we will use the Choi-Jamio{\l}kowski-Kraus-Sudarshan (CJKS) isomorphism \cite{CJKS} to define it. Now however, we define it by a maximization over the density operators on which the relevant maps act. More precisely, we define 
%
%In the above Eq.3 we may define the distance on the space of maps on $ S $ by using the $ CJKS $ representation in the following way. We define
%
%\begin{equation}
%\mathcal{D}(\Lambda||\Lambda^{\prime}) = \mathbf{D} \big(\mathbf{I}\otimes \Lambda(|\Psi^{+}\rangle \langle\Psi^{+}|) || \mathbf{I} \otimes \Lambda^{\prime}(|\Psi^{+}\rangle \langle\Psi^{+}|)\big)
%\end{equation}
%where $ |\Psi^{+}\rangle = \dfrac{1}{\sqrt{d}} \sum_{i=0}^{d-1} |ii\rangle $, and $ \mathbf{I} $ is the identity map on the space of operators on a "reference" Hilbert space $ H_{R} $, which has the same dimension $ d $, as $ H_{S} $.
%
%The distance $ \mathcal{D} $ is derived from the distance measure $ \mathbf{D} $ which is defined on the space of density matrices. Thus,
\begin{equation}
\mathcal{D}(\Lambda || \Lambda^{\prime}) = \max_{\rho} \mathbf{D}(\Lambda(\rho) || \Lambda^{\prime}(\rho))
\label{four}
\end{equation}
where $ \mathbf{D} $ is a distance measure defined on the space of density operators, which forms the domain of the maps involved in $ \mathcal{D}$.  
%{\color{red}See Ref.\cite{channel_distance} in this regard.}\
See Refs. \cite{leditzky} in this regard, which uses the concept of a generalized channel divergence between a channel and a set of channels, and which may also be used to quantify the distance we use in the space of channels.

We therefore have 
\begin{equation}
N^{\epsilon}(\Lambda_{S}) = \min_{\tilde{\Lambda}_{S}^{\epsilon} \in S^{\large \epsilon}} \max_{\rho^{S}_{0}} \mathbf{D} \big( \Lambda_{S}(\rho_{0}^{S})||\tilde{\Lambda}_{S}^{\epsilon}(\rho_{0}^{S}) \big),
\label{five}
\end{equation}
where we have involved ourselves in an additional maximization over all the initial states $ \rho_{0}^{S} $. For a fixed $ \tilde{\Lambda}_{S}^{\epsilon} $, if $ \bar{\rho}^{S}_{0} $ is the state that maximizes $ \mathbf{D}(\Lambda_{S}(\rho_{0}^{S})||\tilde{\Lambda}_{S}^{\epsilon}(\rho_{0}^{S}))$, then
\begin{equation}
\mathbf{D} \big( \Lambda_{S}(\bar{\rho}_{0}^{S})||\tilde{\Lambda}_{S}^{\epsilon}(\bar{\rho}_{0}^{S}) \big)  \leq  \mathbf{D} \big( \Lambda_{SE}(\bar{\rho}^{S}_{0} \otimes \rho^{E}_{0}) || \tilde{\Lambda}^{\epsilon}_{SE}(\bar{\rho}^{S}_{0} \otimes \rho^{E}_{0}) \big) 
\label{six}
\end{equation}
where we have assumed that the distance $ \mathbf{D} $ satisfies the inequality $ \mathbf{D}($tr$_{p}\sigma ||$ tr$_{p}\varrho) \leq \mathbf{D}(\sigma || \varrho)$, where tr$ _{p} $ is partial trace over the system denoted by ``\(p\)".  Examples of such distances are trace-norm, relative entropy, etc. \cite{wehrl, books-qic,wilde1,channel_distance}. 

In the special case $\epsilon=0$, 
%Consequently, 
the minimization in Eq. (\ref{three}) will be over maps that lead to time-evolved states $ \tilde{\rho}^{SE}(t) $ for which $ I_{Q}(\tilde{\rho}^{SE}(t))=0$, $ \forall t, \rho^{S}_{0}, \rho^{E}_{0} $. $ I_{Q}(\rho^{AB})= 0$ implies that the state $ \rho^{AB} $ is a product of individual states of the component systems. A product state for $ SE $ at all times for an initial product state of $ SE $ can appear in the following way.

The evolution of $ SEE_{1} $ is unitary, which can be global (i.e., entangling), and hence, the entanglement and other classical and quantum correlations \cite{HHHH, bera} that arise between $ S $ and $ EE_{1} $ may remain between parts of $ S $ and parts of $ EE_{1} $ or between the wholes ($ S $ and $ EE_{1} $), unless the unitary is very special. However, it may so happen that the interaction between $ S $ and $ E $ is weaker (or equivalently, the information flow between $ S $ and $ E $ is slower) than that between $ E $ and $ E_{1} $, so that any entanglement (or other correlations) created between $ S $ and $ E $ are transferred, and consequently hidden, in entanglement (or other correlations)  between $ S $ and $ E_{1} $. In other words, after an interaction between $ S $ and $ E $, the state of $ E $ at a given time $ t $, is quickly transferred into the recesses of $ E_{1} $, and replaced with a $ \rho_{0}^{E} $, which has no correlations with $ S $, and this is done before the next interaction of $ E $ with $ S $ starts off. 
%This is the Markovian-like limit in our scenario. 
We denote the set of all such reduced dynamical maps by $ S^{0} $, and call them as Markovian-like, as discussed in Section \ref{dui}. 

The description above is close to the collision model \cite{collision}, though having the difference that the interaction of $S$ and $E$ is not unitary, as described in collision model. In this sense, the scenario we consider is a more general picture. Indeed, some assumptions are similar, e.g., $E$ never gets replaced with that part of $E_{1}$ which already shares correlation with $S$, and that $E_{1}$ is infinitely large. In the special case of $\Lambda_{SE}$ being unitary, our model becomes same as collision model.

%\textbf{Let us first consider the special case of $\mathbf{\epsilon=0}$.} 
\section{Behavior of distance based non-Markovianity measure}
\label{pnach}

Next, we want to numerically investigate the behaviour of the distance based measure of non-Markovianity. As discussed in the beginning, calculating $\epsilon$-nonmarkovianity for a non-zero $\epsilon$ is numerically challenging. Nevertheless, in order to get an idea about how a distance based measure will behave temporally, we resort to the usual picture of Markovian and non-Markovian maps in terms of their divisibility. As we will see below, a distance measure can indeed capture the traits of a non-markovian dynamics.

\subsection{Amplitude damping channel} 
\label{pnach-ka}

\begin{figure}[ht]
\includegraphics[width=0.5\textwidth]{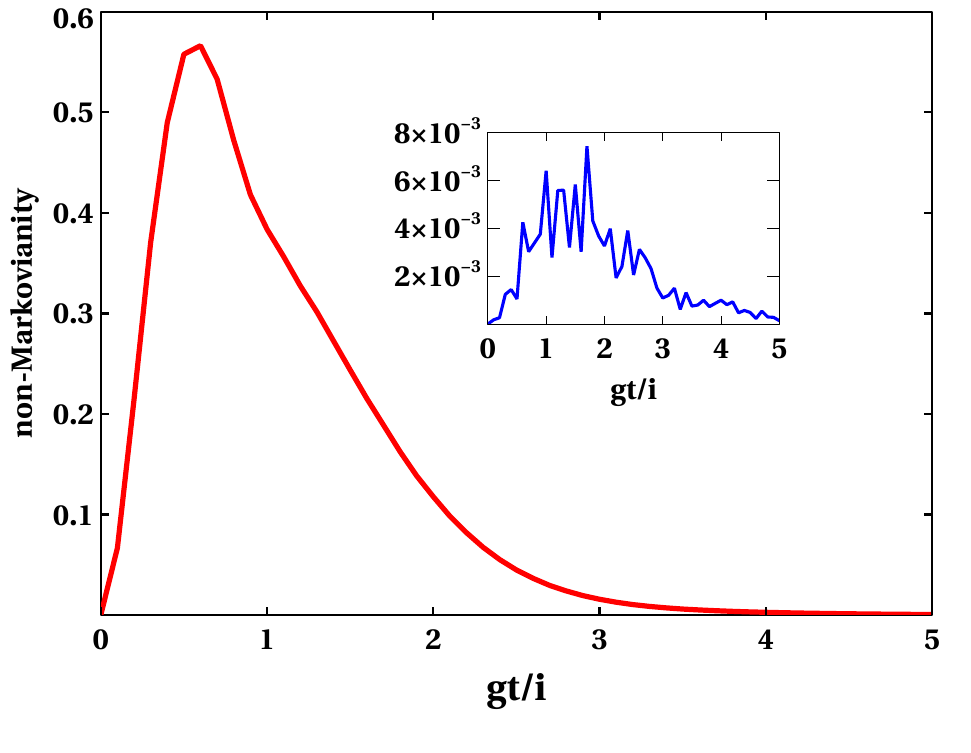}
\caption{(Color online.) Non-Markovianity of the  amplitude damping channel for the ``max-distance'' (red-solid curve) 
%and ``min-distance'' (violet dotted curve)
 approach. We plot the non-Markovianity of the amplitude damping channel for \(\lambda \kappa =4\), 
\(\gamma_0 \kappa = 4\), where \(\kappa\) is a constant having the unit of time. Note that for such choice of \(\lambda \kappa\) and \(\gamma_0 \kappa\), \(g \kappa\) is purely imaginary. 
The horizontal axis represents \(gt/i\), while the vertical one represents non-Markovianity for 
\(\epsilon = 0\). Both axes are dimensionless. The distance between two density matrices is calculated by using the trace distance. %\cite{trace-distance}. 
The Haar-uniform searches are performed over \(2000\) values 
of the pair \(\gamma_0\), and over \(2000\) density matrices, for each value of \(gt/i\). In the inset, the blue solid curve is the non-Markovianity in ``min-distance'' approach (see Section \ref{aaT}), for the same choice of parameters. Note that the increased oscillations in the min distance approach is possibly a numerical artefact, and if we disregard them, the general trend of the curves for min- and max-distance approaches are similar.
%A schematic diagram showing the system $S$ immersed in environment $E$. The joint system $ SE $ is in contact with a larger environment $ E_{1} $.
%We have used the function \(\exp(-A (gt_{MN}/i)) \cos(B (gt_{MN}/i)) + C\), where \(A\), \(B\), and \(C\) are free parameters to plot the curve. The best fit, obtained by using the non-linear least square Marquardt-Levenberg algorithm approach, is given by $A=-4.88198$, $B=-0.00715067$, and $C=0.606672$. The error as quantified by the final sum of squares of residuals is 0.0255422. 
}
\label{jonmo-moder-}
\end{figure}

To exemplify the behavior of non-Markovianity, we consider 
%For the investigation, we have considered 
an amplitude damping channel \cite{books-qic,Myatt,Turchette,Chirolli,victor,Zou,pirandola-old, wilde2,pirandola_new}. The explicit example is taken from Ref. \cite{victor}. The channel has the Lindblad generator $ \mathcal{L}(\rho(t)) $ %(see \cite{victor}) 
given by
%\begin{equation}
\begin{eqnarray}
%\begin{split}
%&&\mathcal{L}(\rho(t))=\mathcal{L}_{1}(\rho(t)) + e^{\beta}\mathcal{L}_{2}(\rho(t)), \nonumber
%\\
%&&\mathcal{L}_{1}(\rho(t))=\sigma_{+}\rho(t)\sigma_{-} - \dfrac{1}{2}\{\sigma_{-}\sigma_{+}, \rho(t)\}, \nonumber
%\\
\mathcal{L}(\rho(t))=\sigma_{-}\rho(t)\sigma_{+} - \dfrac{1}{2}\{\sigma_{+}\sigma_{-}, \rho(t)\},
%\end{split}
\end{eqnarray}
where $ \sigma_{-} $  and $ \sigma_{+} $ are the qubit raising and lowering operators, and we are 
in the limit of a zero temperature bath.
%$ \beta $ is the inverse temperature of the bath. 
The corresponding master equation is $ \dot{\rho}=\gamma_{a}(t)\mathcal{L}(\rho(t)) $, where we choose the time-dependent ``decay rate'' $ \gamma_{a}(t) $ as \cite{victor}
%is given by
\begin{equation}
\gamma_{a}(t)=\dfrac{2\lambda\gamma_{0} \sinh\ \dfrac{tg}{2}}{g \cosh\ \dfrac{tg}{2} + \lambda \sinh\ \dfrac{tg}{2}},~~  g=\sqrt{\lambda^{2}-2\gamma_{0}\lambda}.
\label{seven0}
\end{equation}
Here, \(\gamma_0\) and \(\lambda\) are two system parameters that can be used to decide about the Markovianity in terms of divisibility of the dynamics, as shown by Mukherjee \emph{et al.} \cite{victor}. 
%Depending on the values of the parameters $ \{\gamma_{0}, \lambda \} $, the bath can behave as Markovian-like or non-Markovian. 
When $ \lambda > 2\gamma_{0} $, the evolution is divisible and when $ \gamma_{0} > \lambda /2 $, the evolution is non-divisible. 
The non-divisible map $ \Lambda_{S} $ is constructed by choosing a particular pair of values of $ \gamma_{0} $ and $ \lambda $ such that $ \gamma_{0} > \lambda /2 $.
%and a particular value of time $ t_{NM} $.
 Keeping the value of $ \lambda $ the same as that for the non-divisible map, we generate a class of divisible maps $ \tilde{\Lambda}_{S}^{0} $, by randomly choosing $ \gamma_{0} $,
 %and time $ t_{M} $, 
 from a uniform distribution, while satisfying the condition $ \lambda > 2\gamma_{0} $.  In parallel, we also generate the set of all density operators $ \rho^{S}_{0} $ on \(\mathbb{C}^2\) by Haar-uniformly generating pure states on the larger Hilbert space \(\mathbb{C}^2 \otimes \mathbb{C}^2\). The generation of the density matrices is therefore according to the ``induced metric'' \cite{karol, induced}. We now apply both the maps on the elements of this set of density operators to obtain the time evolved states $ \Lambda_{S}(\rho^{S}_{0}) $ and $ \tilde{\Lambda}_{S}^{0}(\rho^{S}_{0}) $, and maximize the trace distance  between them over this set of density operators. 
 The ``trace distance'' between two density matrices \(\varrho\) and \(\varsigma\) 
is given by 
\(\mbox{tr}\sqrt{(\varrho - \varsigma)(\varrho - \varsigma)^\dagger}\).
 The trace distance thus obtained, is further minimized over the set of divisible maps generated by varying \(\gamma_0\) for a fixed \(\lambda\), with $ \lambda > 2\gamma_{0} $. 
%For this, we randomly choose $ \{\gamma_{0}, t_{M} \} $ from uniform distribution a large number of times, and minimize the trace distance w.r.t this set of $ \{\gamma_{0}, t_{M} \} $. 
The entire optimization process is executed for 
%by varying 
different points on the $ t $-axis. The data is presented in Fig. \ref{jonmo-moder-} as the red solid curve. The non-Markovianity increases initially, starting to decrease after a certain time and then decaying to zero, i.e. the map becomes Markovian after a time.
It is important to mention here that the optimization is performed under the assumption that 
the optimal Markovian-like map for a given non-divisible amplitude damping channel is attained 
within the class of divisible  amplitude damping channels. 
%The obtained values therefore form, 
%in the worst-case scenario, upper bounds of the actual values.

It is of course true that time-dependence in the parameters of the dynamical equation are not necessary for the optimization process, as only the outputs of the non-divisible channel and the nearest divisible channel are used in the definition of the non-Markovianity measure. But this is true if we are able to consider the optimization over all Markovian-like maps. This we have been unable to perform. We have therefore considered the time-dependent decay rates for all times for optimization process. For every time, it provides another element of the space of divisible maps, and gives us a chance of having a better estimate of the non-Markovianity measure. 

%We thank the Referee for this comment. 
While the optimization is carried out under the assumption that the optimal Markovian-like map will be obtained from within the class from which the non-divisible map is chosen, 
%
%this is solely to has led to this simplification and it is within this assumption that we have been able to calculate the values of the measure for the examples considered. These calculations are given to provide the Reader with a feel of how the measure behaves for well-known maps. Please note however that 
the ``\(\epsilon\)-entanglement bound'' on the \(\epsilon\)-nonmarkovianity measure, that we obtain below, is independent of these assumptions.

\subsection{Phase damping channel}
\label{pnach-kha}

\begin{figure}[t]
\includegraphics[width=0.5\textwidth]{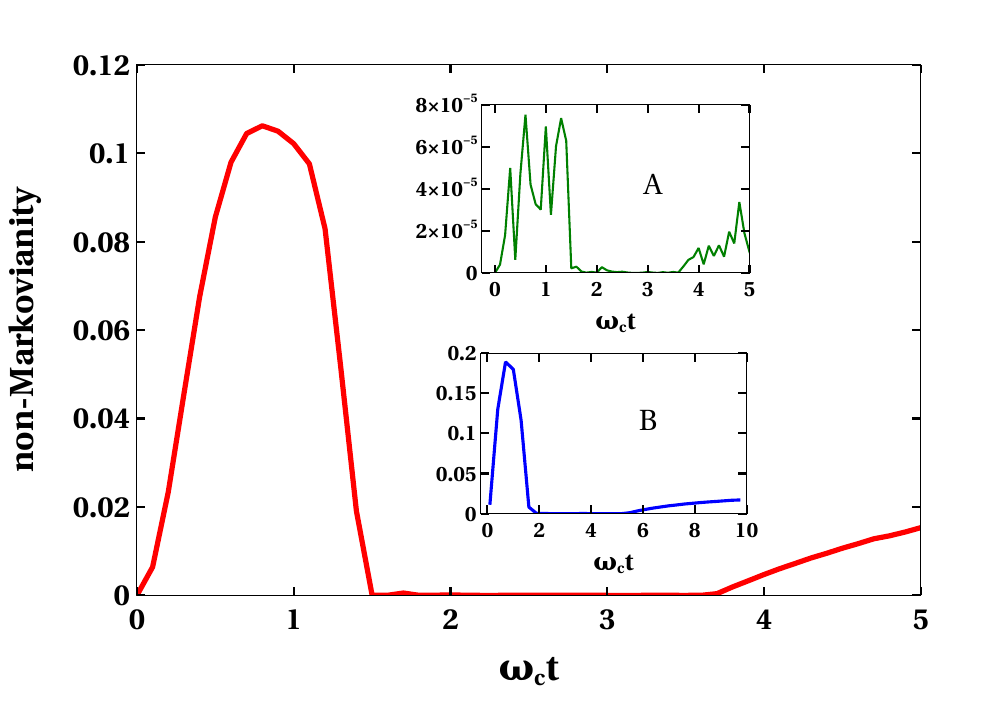}
\caption{(Color online.) Non-Markovianity of the phase damping channel with respect to time $ t $ for ``max-distance'' (red-solid curve) approach, where $ s_{NM}=2.5 $. The non-Markovianity decays, remains zero for finite interval, and then revives again. Inset:(A) Non-markovianity of the phase damping channel for ``min-distance'' approach (see Section \ref{aaT}) when $ s_{NM}=2.5 $. (B) Non-Markovianity of the phase damping channel in ``max-distance'' approach when $ s_{NM}=2.8 $. Both the decay and revival occurs at later times compared to the case when $ s_{NM}=2.5 $.
%We plot the non-Markovianity of the generalized amplitude damping channel for \(\lambda \kappa =4\), 
%\(\gamma_0 \kappa = 10\), where \(\kappa\) is a constant having the unit of time. Note that for these choice of \(\lambda \kappa\) and \(\gamma_0 \kappa\), \(g \kappa\) is purely imaginary. 
%The horizontal axis represents \(gt_{NM}/i\), while the vertical one represents non-Markovianity for 
%\(\epsilon = 0\). Both axes are dimensionless. The Haar-uniform searches are performed over \# values 
%of the pair \((\{\gamma_0, t_M\}\), and over \# density matrices, for each value of \(gt_{NM}/i\). 
%A schematic diagram showing the system $S$ immersed in environment $E$. The joint system $ SE $ is in contact with a larger environment $ E_{1} $. 
}
\label{ph-damp}
\end{figure}

Besides the amplitude damping channel, we also want to investigate how $ N^{\epsilon}(\Lambda_{S}) $ behaves during evolution through a phase damping channel \cite{books-qic,Luczka, Palma, HJM}. The explicit form of the channel is taken from Ref. \cite{HJM}. The master equation of the phase damping channel is 
\begin{equation}
\dfrac{d\rho}{dt}=\frac{1}{2}\gamma(t)[\sigma_{z}\rho(t)\sigma_{z} - \rho],
\end{equation}
where $ \sigma_{z} $ is the Pauli $ z $-matrix. We choose the time-dependent ``dephasing rate'', $ \gamma(t) $, as \cite{HJM}
%is given by
\begin{equation}
\gamma(t)= \int d\omega J(\omega) \coth(\hbar \omega/2 k_{B} T) \sin(\omega t)/\omega,
\end{equation}
where the integration is over the frequency of the bath-modes denoted by $ \omega $, $ J(\omega) $ is the ``spectral function'' of the bath. 
%and we have assumed $ \hbar=1 $ .
If we take our bath of consideration to have Ohmic-like spectra, then the spectral function is
\begin{equation}
J(\omega)=\frac{\omega^{s}}{\omega_{c}^{s-1}}e^{-\omega/\omega_{c}},
\end{equation}
where $ \omega_{c} $ is the ``frequency cut-off'', and $ s $ is the ``Ohmicity parameter'', which decides whether the bath will be Ohmic ($ s=1 $), sub-Ohmic ($ 0<s<1 $), or super-Ohmic ($ s>1 $). Keeping attention at the absolute zero temperature limit, we get the following expression for the dephasing rate:
\begin{equation}
\gamma_{\bar{0}}(t)=\omega_{c}[1+(\omega_{c}t)^{2}]^{-s/2}\Gamma(s)\sin[s\arctan(\omega_{c}t)],
\end{equation}
where \(\Gamma(s)\) denotes the Gamma function, given by 
\(\Gamma(s) = \int_0^{\infty} x^{s-1} \exp{(-x)} dx\). While the Gamma function can be generalized to more general values of \(s\), we will only use it for real \(s>0\).
In \cite{HJM}, Haikka \emph{et al}. have demonstrated that the non-divisibility of this channel  is observed only when $ s>2 $. To numerically study $ N^{\epsilon}(\Lambda_{S}) $ for a non-divisible map, we take a particular value of $ s = s_{NM} $, such that $ s_{NM}>2 $, corresponding to the map $ \Lambda_{S} $. We generate $2000$ values of $ s = s_{M} $ in the interval $ 0<s<2 $, each corresponding to a divisible map, and for each $ s_{M} $, we Haar-uniformly generate $2000$ random density matrices. 
The optimized $ N^{\epsilon}(\Lambda_{S}) $ has been presented in Fig. \ref{ph-damp} with respect to the $ \omega_{c}t $ along the horizontal axis. Similar to the amplitude damping channel, the non-Markovianity increases at first, reaches a maximum, then decays to zero. Notably, it revives again from zero after a time-interval. This non-monotonic behaviour of several physical quantities associated to the system is typical of a non-Markovian dynamics.

%In Fig. \#\#, we have numerically studied the behaviour of non-Markovianity, as 
%quantified by \(N^{\epsilon}(\Lambda_{S})\), 
%%optimized distance $ \mathbf{D} \big( \Lambda_{S}(\rho_{0}^{S})||\tilde{\Lambda}_{S}^{0}(\rho_{0}^{S}) %\big) $ on the right-hand-side of eq. (\ref{four}) 
%as a function of  time, for the special case of $ \epsilon=0 $. 

\section{Entanglement-based bound on non-Markovianity}
\label{chhoi}

\begin{figure}[t]
\includegraphics[width=0.5\textwidth]{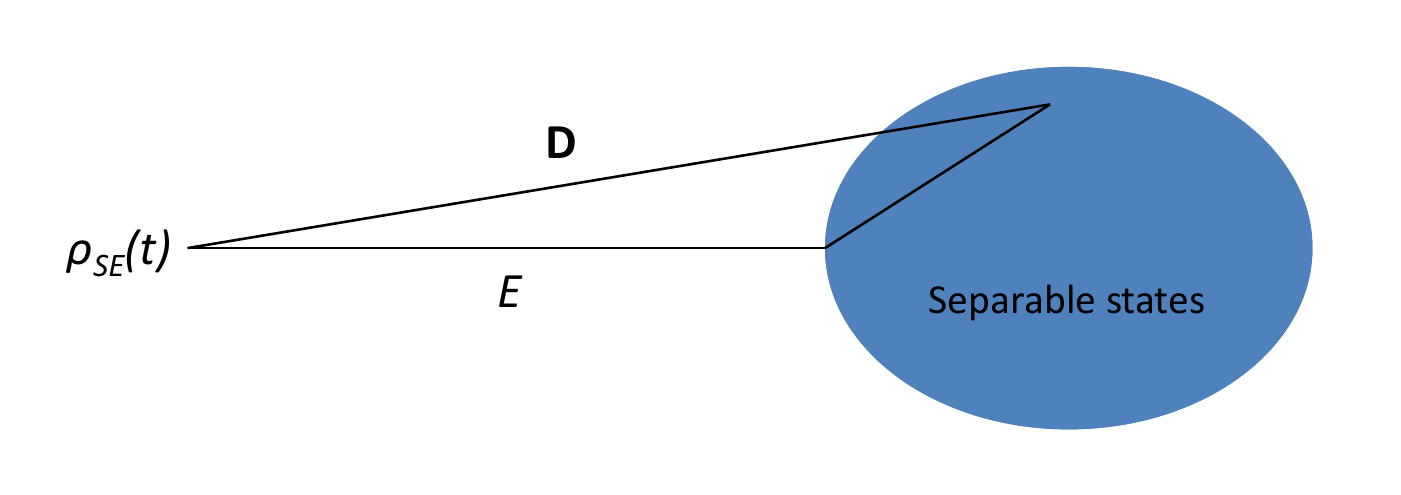}
\caption{(Color online.) Geometric representation of the inequality between non-Markovianity and system-environment entanglement. We consider the case of $ \epsilon=0$. The shaded region represents the convex set of states that are separable in the system-environment
bipartition. The optimal state that attains the optimization in the definition of non-Markovianity is 
on or inside the set of separable states. The distance of this state to the time-evolved state 
\(\rho_{SE}(t)\)  must be greater than the sum of the other two sides of the triangle depicted in the figure. These latter sides however are respectively a distance-based entanglement of the evolved state and a bound on the diameter of the set of separable states. We have assumed that the distance measure on the space of density matrices satisfies the triangle inequality. 
%(In the inset)(A) The non-Markovianity measure in ``min-distance'' approach for the same choice of parameters as in the main plot. (B) The non-Markovianity measure in ``max-distance'' approach when $ s = 2.8 $, all other parameters remaining same. 
%ball of time-evolved states when $ \epsilon=0$, the distance $\mathcal{D}$ of time-evolved state $ \rho_{SE}(t) $ from a point on the separable ball, the relative entropy of entanglement $ E_{R} $ of the time-evolved state $ \rho_{SE}(t) $.
}
\label{schematic2}
\end{figure}

Going back to the scenario of general channels, but still remaining with the case when $ \epsilon=0$, we have 
%so $ \tilde{\Lambda}^{0}_{SE}(\bar{\rho}^{S}_{0} \otimes \rho^{E}_{0}) = \bar{\rho}^{S}(t) \otimes \rho^{E}(t)$, i.e at any time $ t $ the joint system $ SE $ is in a product state. In this case,
\begin{equation}
\mathbf{D} \big( \Lambda_{SE}(\bar{\rho}^{S}_{0} \otimes \rho^{E}_{0}) || \tilde{\Lambda}^{0}_{SE}(\bar{\rho}^{S}_{0} \otimes \rho^{E}_{0}) \big) \geq E \big(\Lambda_{SE}(\bar{\rho}^{S}_{0} \otimes \rho^{E}_{0}) \big),
\label{seven}
\end{equation}
where $ E $ is a distance-based entanglement defined as the minimum distance of a state from the set of separable states \cite{HHHH}. The inequality (\ref{seven}) holds by virtue of the fact that $ \tilde{\Lambda}_{SE}^{0}(\bar{\rho}_{0}^{S} \otimes \rho_{0}^{E}) $ is a separable and indeed a product state. In case the distance $ \mathbf{D} $ is the relative entropy on the space of density operators, $ E $ is the relative entropy of entanglement \cite{VPRK, vedral} of its argument. Let us now assume that $ \mathbf{D} $ satisfies the triangle inequality, which is in fact not satisfied by relative entropy distance. We then obtain
\begin{equation}
\mathbf{D} \big( \Lambda_{SE}(\bar{\rho}^{S}_{0} \otimes \rho^{E}_{0}) || \tilde{\Lambda}^{0}_{SE}(\bar{\rho}^{S}_{0} \otimes \rho^{E}_{0}) \big) \leq E \big(\Lambda_{SE}(\bar{\rho}^{S}_{0} \otimes \rho^{E}_{0}) \big) + d,
\label{eight}
\end{equation}
where $ d $ is the ``diameter" of the convex set of separable states \cite{ZHL}. The diameter, $ d $, of a set  $ \mathcal{S} $, is defined as
\begin{equation}
d(\mathcal{S}) = \max_{\rho, \sigma \in \mathcal{S}} \mathbf{D}(\rho, \sigma)
\label{nine}
\end{equation}
A geometric representation of the relation (\ref{eight}) is given in Fig. \ref{schematic2}. 
%We, therefore have, 
By combining relations (\ref{six}) and (\ref{eight}) with definitions (\ref{three}) and (\ref{five}),
we have
\begin{equation}
N^{0}(\Lambda_{S}) \leq E \big(\Lambda_{SE}(\bar{\rho}^{S}_{0} \otimes \rho^{E}_{0}) \big) + d.
\label{ten}
\end{equation}
This relation is true for all extensions of $ \Lambda_{S} $ into $ \Lambda_{SE} $ and for all $ \rho_{0}^{E} $. Consequently,
\begin{equation}
N^{0}(\Lambda_{S}) \leq \min_{\rho_{0}^{E}, \Lambda_{SE}}[E \big(\Lambda_{SE}(\bar{\rho}^{S}_{0} \otimes \rho^{E}_{0}) \big) + d],
\label{eleven}
\end{equation}
where the minimization is over all $ \rho_{0}^{E} $ and over all extensions of $ \Lambda_{S} $ into $ \Lambda_{SE} $. Note that the diameter $ d $ is inside the optimization process, and not independent of it.

%\min_{\tilde{\rho}^{SE}(t)} \mathcal{D}()

\section{\(\epsilon\)-entanglement and \(\epsilon\)-nonmarkovianity}
\label{saat}

We now consider the general case, i.e., when $ \mathbf{\epsilon \neq 0 }$. In this case, the time evolved state $ \tilde{\rho}^{SE}(t) = \tilde{\Lambda}^{\epsilon}_{SE}(\bar{\rho}^{S}_{0} \otimes \rho^{E}_{0}) $ is no longer a product state, and instead it satisfies the inequality $ I(\tilde{\Lambda}^{\epsilon}_{SE}(\bar{\rho}^{S}_{0} \otimes \rho^{E}_{0})) \leq \epsilon $, a weaker condition, that does not require the argument to be product (for \(\epsilon \ne 0\)). The set of all states \(\eta^{SE}\) that satisfy 
\(I_Q(\eta^{SE}) \leq \epsilon\)
does not form a convex set; however the set of convex combinations of all such states is of course a convex set and we call this set as the set of \emph{$ \epsilon$-separable} states. For $\epsilon = 0$, this set becomes the set of convex combinations of product states, i.e. the set of separable states which have non-zero classical correlation. However, to calculate $\epsilon$-nonmarkovianity in the case $\epsilon=0$, we still need to minimize the distance function from the set of product states itself. In the present case of $\epsilon\neq 0$, the minimum distance possible between $ \Lambda_{SE}(\bar{\rho}^{S}_{0} \otimes \rho^{E}_{0}) $ and the set of $ \epsilon$-separable states, is referred to as 
\emph{$ \epsilon$-entanglement}, $E^{\epsilon}$, of the state $ \tilde{\rho}^{SE}(t) $. 
In case we use the relative entropy as a measure of distance, this is denoted by $ E_{R}^{\epsilon} $. 

We take a short break here from the main stream of the paper, and prove that the entanglement, as quantified by the relative entropy of entanglement \cite{VPRK, vedral} (denoted by \(E_R\)), of 
an arbitrary two-party quantum state whose \(\epsilon\)-entanglement is vanishing, cannot be higher than \(\epsilon\). This, we believe, would go some way in legitimizing the use of the term ``\(\epsilon\)-entanglement, as for small \(\epsilon\), only weakly entangled states would have zero \(\epsilon\)-entanglement. Let us consider the two-party quantum state \(\rho_{AB}\) such that \(E_R^\epsilon(\rho_{AB}) = 0\). Therefore, by definition, \(I_Q (\rho_{AB}) \leq \epsilon\). Now, as we have already mentioned, the quantum mutual information of a bipartite state is the relative entropy distance of it from the tensor product of its local parts, and the latter is a separable state. Therefore, by definition of the relative entropy of entanglement, \(E_R (\rho_{AB}) \leq I_Q(\rho_{AB})\), so that 
\(E_R(\rho_{AB}) \leq \epsilon\).

Going back to where we left before the last paragraph, if $ d_{\epsilon} $ is the diameter of the set of \(\epsilon\)-separable states, we get the following inequality:
\begin{equation}
\mathbf{D} \big( \Lambda_{SE}(\bar{\rho}^{S}_{0} \otimes \rho^{E}_{0}) || \tilde{\Lambda}^{\epsilon}_{SE}(\bar{\rho}^{S}_{0} \otimes \rho^{E}_{0}) \big) \leq E^{\epsilon} \big(\Lambda_{SE}(\bar{\rho}^{S}_{0} \otimes \rho^{E}_{0}) \big) + d_{\epsilon}.
\label{twelve}
\end{equation}
In this case, the relation (\ref{eleven}) is replaced by
\begin{equation}
N^{\epsilon}(\Lambda_{S}) \leq \min_{\rho_{0}^{E}, \Lambda_{SE}} [E^{\epsilon} \big(\Lambda_{SE}(\bar{\rho}^{S}_{0} \otimes \rho^{E}_{0}) \big) + d_{\epsilon}].
\label{thirteen}
\end{equation}

\section{The min-distance} 
\label{aaT}

The measures of non-Markovianity and \(\epsilon\)-nonmarkovianity depended, among other things, on the fact that we perform a maximization over the set of density matrices on the system \(S\). See Eq. (\ref{four}). 
%, instead of a maximization over all density operators, 
Let us refer to this strategy as that of ``max-distance''. 
This however is hardly a unique strategy, and in particular, one can certainly define 
the distance between the maps by using a \emph{minimization} over the density operators, i.e., by using the 
distance
\begin{equation}
\mathcal{D}_m(\Lambda || \Lambda^{\prime}) = \min_{\rho} \mathbf{D}(\Lambda(\rho) || \Lambda^{\prime}(\rho)).
\label{fourteen}
\end{equation}
So, the definition of non-Markovianity accordingly changes to
\begin{equation}
N^{\epsilon}_m(\Lambda_{S}) = \min_{\tilde{\Lambda}_{S}^{\epsilon} \in S^{\large \epsilon}} \min_{\rho^{S}_{0}} \mathbf{D} \big( \Lambda_{S}(\rho_{0}^{S})||\tilde{\Lambda}_{S}^{\epsilon}(\rho_{0}^{S}) \big).
\label{fifteen}
\end{equation}
We refer to this approach as that of the ``min-distance''.
Suppose that for a fixed $ \tilde{\Lambda}_{S}^{\epsilon} $, $\bar{\bar\rho}_{0}^{S}$ is the state that minimizes $ \mathbf{D}(\Lambda_{S}(\rho_{0}^{S})||\tilde{\Lambda}_{S}^{\epsilon}(\rho_{0}^{S}))$. Proceeding as in the case of ``max-distance'', 
we can see that \textbf{in the special case of} $\mathbf{\epsilon=0} $, the relation (\ref{eight}) changes to
\begin{equation}
\mathbf{D} \big( \Lambda_{SE}(\bar{\bar\rho}^{S}_{0} \otimes \rho^{E}_{0}) || \tilde{\Lambda}^{0}_{SE}(\bar{\bar\rho}^{S}_{0} \otimes \rho^{E}_{0}) \big) \leq E \big(\Lambda_{SE}(\bar{\bar\rho}^{S}_{0} \otimes \rho^{E}_{0}) \big) + d.
\label{sixteen}
\end{equation}
Accordingly, the relation (\ref{eleven}) changes to
\begin{equation}
N^{0}_m(\Lambda_{S}) \leq \min_{\rho_{0}^{E}, \Lambda_{SE}}[E \big(\Lambda_{SE}(\bar{\bar\rho}^{S}_{0} \otimes \rho^{E}_{0}) \big) + d],
\label{seventeen}
\end{equation}
where the entanglement function is the same as before, while the actual quantity has changed to $ E \big(\Lambda_{SE}(\bar{\bar\rho}^{S}_{0} \otimes \rho^{E}_{0}) \big) $,
%in R.H.S of the inequality is different than the entanglement 
in contrast to $ E \big(\Lambda_{SE}(\bar{\rho}^{S}_{0} \otimes \rho^{E}_{0}) \big) $ in the ``max-distance'' case.

The \(\epsilon \ne 0\) can be similarly derived.
It is important to stress here that despite the similarity of notation and the algebra, we have here a 
completely independent bound on an independent measure of non-Markovianity of a general dynamical map $ \Lambda_{S} $, as compared to the case of ``max-distance''.

Similar to the case of ``max-distance'', here also, we have numerically studied the behavior of non-Markovianity for the amplitude damping channel. This is presented in the inset of Fig. \ref{jonmo-moder-}. The corresponding diagram for the phase damping channel is an inset of Fig. \ref{ph-damp}. The value of the optimized distance is very small ($\sim 10^{-3}$-$10^{-5}$). The size of the set used for optimization is not large enough in this case for the first significant digit to converge, despite the simulation taking long time. This leads to the oscillatory behaviour observed in the figures corresponding to ``min-distance". Nevertheless the envelope of the oscillations has an overall same behavior as that observed in the max-distance approach.
%shows the behaviour of the quantity on the R.H.S of eq. () with time $ t_{NM} $.

\begin{figure}[t]
\includegraphics[width=0.45\textwidth]{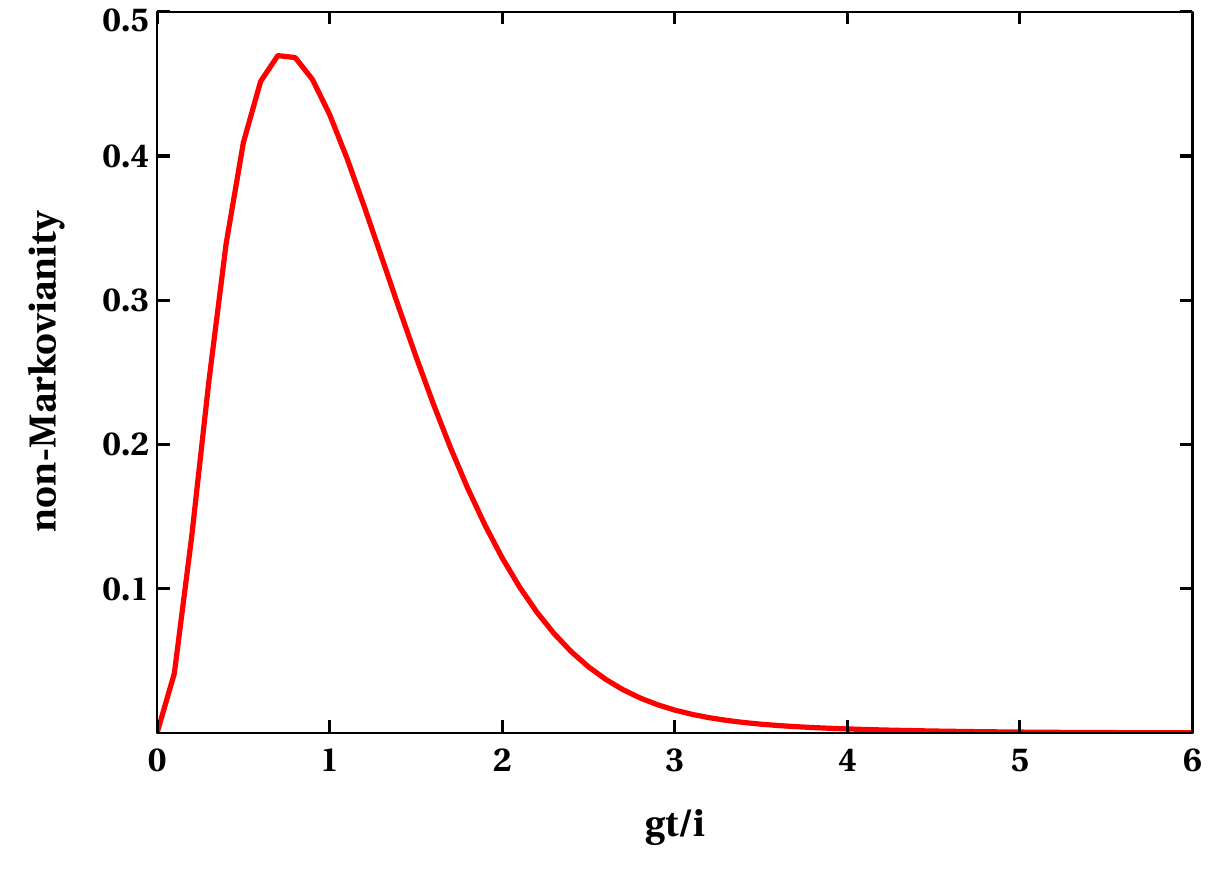}
\caption{(Color online.) Non-Markovianity of the amplitude damping channel in the CJKS approach for  \(\lambda \kappa =4\), \(\gamma_0 \kappa = 10\),
The notations in this figure remain the same as in Fig. \ref{jonmo-moder-}, except that the non-Markovianity is defined here from the CJKS approach, and 
that the curve is plotted by simply joining the data points.
%We plot the non-Markovianity of the generalized amplitude damping channel for \(\lambda \kappa =4\), 
%\(\gamma_0 \kappa = 10\), where \(\kappa\) is a constant having the unit of time. Note that for these choice of \(\lambda \kappa\) and \(\gamma_0 \kappa\), \(g \kappa\) is purely imaginary. 
%The horizontal axis represents \(gt_{NM}/i\), while the vertical one represents non-Markovianity for 
%\(\epsilon = 0\). Both axes are dimensionless. The Haar-uniform searches are performed over \# values 
%of the pair \((\{\gamma_0, t_M\}\), and over \# density matrices, for each value of \(gt_{NM}/i\). 
%A schematic diagram showing the system $S$ immersed in environment $E$. The joint system $ SE $ is in contact with a larger environment $ E_{1} $. 
}
\label{-sakal-}
\end{figure}

\section{The CJKS approach to non-Markovianity} 
\label{noi}

In the analysis until now, we have defined distance on the space of maps by using a distance on the space of density operators on which the maps act and a corresponding double optimization. See Eqs. (\ref{four}) and (\ref{fourteen}). However, instead of using these approaches (viz., the ``max-distance'' and the ``min-distance'' ones), in Eq. (\ref{three}), we may define the distance on the space of maps on $ S $ by using the %Choi-Jamio{\l}kowski-Kraus-Sudarshan 
CJKS  representation  in the following way. We define
\begin{equation}
\mathcal{D}(\Lambda||\Lambda^{\prime})=\mathbf{D}(\mathbf{I}\otimes \Lambda(|\Psi^{+}\rangle\langle \Psi^{+}|)||\mathbf{I}\otimes \Lambda^{\prime}(|\Psi^{+}\rangle\langle \Psi^{+}|))
\label{eighteen}
\end{equation}
where $ |\Psi^{+}\rangle=\dfrac{1}{\overline{d}^{\frac{1}{2}}} \sum_{i=0}^{\overline{d}-1} |ii\rangle$, and $ \mathbf{I} $ is the identity map on the space of operators on a ``reference" Hilbert space, $ H_{R} $, which has the same dimension $ \overline{d} $, as $ H_{S} $. Note that \(|\Psi^{+}\rangle\) is an element of \(H_R \otimes H_S\).
This approach inherits the properties of the CJKS representation, and in particular has the benefit of 
a reduced level of optimization, as compared to the preceding approach. In Fig. \ref{-sakal-}, we provide a 
numerical calculation to exemplify the behavior of the non-Markovianity when seen through this approach, for the amplitude damping channel.

\section{Conclusion}
\label{dos}

To conclude, we have considered a measure of non-Markovianity of a dynamical map on an open quantum system based on the distance of the dynamical map on the reduced system from the set of all ``Markovian-like'' dynamical maps on the same system. We found a quantitative relation between the measure, and the entanglement between the reduced system and the environment. This relation can be used to estimate one of the quantities if we are able to find the other. To exemplify the notion and the relation, we have studied 
amplitude damping and phase damping channels.

Along with considering the case when the maps are exactly Markovian-like, we have also considered the situation when a map violates the  Markovian-like property but only a ``little''. This is quantified by introducing the concept of 
\(\epsilon\)-Markovianity. Correspondingly, we present the notion of being outside this set, and quantify it by proposing the concept of \(\epsilon\)-nonmarkovianity. We then went on to generalize the entanglement-based bound on non-Markovianity mentioned above to the \(\epsilon \ne 0\) case, by suggesting a notion of \(\epsilon\)-separability and \(\epsilon\)-entanglement.

% distance as a function of time. 
%  The total system is initially in a product state $ 
% \rho=\rho_{1}\otimes\rho_{2} $. The aim is to cool qubit 1. This is done by 
% introducing an interaction between the qubits which is given by 
%The qubits interact via the following interacting hamiltonian $H_{int}=g(\Ket{101}\Bra{010}+\Ket{010}\Bra{101})$. The interaction strength $g$ is taken weak enough compared to the 

%\begin{equation}
% H=\sum_{i=1}^{3} E_{i} |1\rangle_i\langle1|+ g(\Ket{101}\Bra{010}+\Ket{010}\Bra{101}).
%\end{equation}

%As the qubits are coupled with heat baths at each time step there is finite probability that it will thermalize. Suppose $ p_{i} $ 

%\begin{equation}
%\label{master}
 %\frac{\partial \rho}{\partial t}  = -i[H_{0}+H_{int},\rho]+\sum_{i=1}^{3}  
%p_{i} (\tau_{i}\otimes Tr_{i}\rho - \rho).
%\end{equation}
%It is necessary to mention that this master equation is valid only in the perturbative regime where $p_i,g<<E_i$ and $p_i<<1$. The thermalization of more than one qubit simultaneously
%is of second order in $p_i$'s an hence can be neglected. The steady state refrigeration with the aforementioned model has been demonstrated in great detail in \cite{popescu10}.
%The refrigeration in the transient regime with the same model is discussed very recently in \cite{brask15}. The methods for solving the master equation is discussed 
%in great detail in \cite{brask15}. 

\begin{acknowledgments}
We acknowledge useful discussions with Aditi Sen(De). We thank Pedro Figueroa-Romero, Stefano Pirandola, and Mark Wilde for useful comments. The research of SD was supported in part by the INFOSYS scholarship for senior students.  SB thanks SERB (DST), Government of India for financial support. The work is supported by the  ``QUEST'' grant of the Department of Science and Technology  of the Government of India.
\end{acknowledgments}

 %\bibliography{refri}
  
\end{document}